\documentclass[prb,superscriptaddress,showpacs,showkeys,twocolumn,10pt]{revtex4-1}
\usepackage{amsfonts,amsmath,amssymb,graphicx,epstopdf,verbatim,color}

\usepackage[english]{babel}

\begin{document}

\title{Many-body effects of a two-dimensional electron gas on trion-polaritons}
\author{Maarten Baeten}
\affiliation{Theory Of Quantum and Complex Systems, Universiteit Antwerpen, Universiteitsplein 1, B-2610 Antwerpen, Belgium}
\author{Michiel Wouters}
\affiliation{Theory Of Quantum and Complex Systems, Universiteit Antwerpen, Universiteitsplein 1, B-2610 Antwerpen, Belgium}
\date{\today}

\begin{abstract}
We theoretically investigate the trion-polariton and the effects of a two-dimensional electron gas on its single particle properties. Focussing on the trion and exciton transitions, we set up an effective model and calculate the optical absorption of the quantum well containing the 2DEG. Including the light-matter coupling, we compute the Rabi splitting and polariton lineshapes as a function of 2DEG density. The role of finite temperature is  investigated. The spatial extent of the trion-polariton is also calculated. We find a substantial charge build-up at short distances as long as the Rabi frequency does not exceed the trion binding energy. All our calculations take into account the Fermi-edge singularity and the Anderson orthogonality catastrophe.
\end{abstract}

\maketitle

\section{Introduction}\label{sec:Introduction}
Embedding a quantum well (QW) inside a semiconductor microcavity results in the formation of the so-called microcavity polaritons; a coherent superpositon of the cavity mode and an exciton, when the cavity is tuned to an excitonic transition in the QW. Over the last 15 years, these bosonic quasiparticles were shown to have many nice properties for studying quantum optics and many-body physics in integrated photonic structures. For example, their photonic component allows for a direct observation of the polaritons and a large coherence length, while the matter component is responsible for polariton interactions.
Bose-Einstein condensation of polaritons and their superfluid properties have been investigated in detail both experimentally and theoretically \cite{iac_review}.

The presence of charges inside the quantum well that could interact with the polaritons, has been put aside until recently. From the theoretical side, the mixed electronic-polaritonic system was proposed to reach superconductivity at higher temperatures, possibly even room temperature, thanks to a strongly attractive electron-electron interaction mediated by the polaritons \cite{laussy}. 

A pioneering experiment with high density modulation doping was performed by Gabbay {\em et al.} \cite{GabbayCohen}, where it is shown that strong light-matter coupling is possible, even though neither an excitonic nor a trionic state could be resolved. Recently, in the experiments by Smolka et al. \cite{atac} polariton formation was observed over a range of electron densities, both with and without a magnetic field.

For what concerns the effect of the electron gas on the polaritons, various contributions both from the experimental \cite{Lagoudakis,Perrin,Das} and theoretical \cite{Kavokin} side have been made. Those works considered the electron-polariton scattering physics, but it was assumed that the polariton quasiparticle itself was not affected by the electron gas. 

The optical excitation of an electron gas (metal or doped semiconductor) is an important problem in many-body physics that was introduced by Mahan \cite{mahan} and to which Nozi\`eres, De Dominicis \cite{nozieres} and Anderson \cite{Anderson} made seminal contributions. Nowadays, the Fermi edge singularity is still an active topic of research that has recently attracted the attention of the community working on ultracold atomic gases \cite{knap,demler,Schirotzek}.

The role  of the electrons on the properties of the exciton-polariton itself for a highly doped quantum well has been investigated in Refs. \onlinecite{glazov,Averkiev}. In our previous work\cite{BW}, we have discussed the single particle exciton-polariton properties, properly taking into account the many-body physics involving the Anderson orthogonality catastrophe (AOC) and Fermi edge singularity (FES). Unfortunately, an exact evaluation of the optical response was only possible for negligible electron-electron interactions. In the low density regime, where trion correlations dominate, this makes it a poor approximation. 
In the case of a low density electron gas, trion-polaritons have been observed and discussed theoretically \cite{Rapaport,Bloch,atac}. A theoretical model for trion-polaritons, that takes into account the AOC and FES, is however still missing. It is the purpose of this paper to fill this gap. We will focus on the linear properties of trion-polaritons, deferring the study of polariton-polariton interactions to a future work.

This paper is organized as follows: in Section \ref{sec:One} the notations used in this article are introduced and we set up our model and show how to calculate the optical response of the two-dimensional electron gas. Section \ref{sec:TP} presents some linear polariton properties such as the polariton energies, lineshapes. In the same Section we calculate the spatial distribution of electrons in the lower polariton state. The last section \ref{ConclusionAndOutlook} gives a summary and outlook. 

\section{Hamiltonian and optical response of the quantum well}\label{sec:One}
\subsection{Model set-up and formalism}

The system under consideration is a quantum well containing a non-interacting two dimensional electron gas (2DEG), placed inside a planar microcavity. For the sake of simplicity we consider a spin polarized 2DEG (the generalization of our results for an unpolarized electron gas is straightforward). When a photon impinges onto an empty QW, an interband electron-hole pair can be created and forms a bound electron-hole pair, the exciton. 
In the presence of a 2DEG, the exciton will interact with the electrons, with the possibility of forming a trionic bound state, where the valence band hole is bound to two electrons. Typically, the trion binding energy is much weaker than the exciton binding energy. Therefore, the trion can be approximated as an electron that is bound to an exciton through an effective electron-exciton potential $V^{X-e}$. 
This approximation reduces the trion physics from a three-body to a two-body problem. Within this approximation, it becomes feasible to treat the interaction of the exciton with all the electrons in the 2DEG. Provided that  the electron-electron interactions can be neglected and the exciton mass tends to infinity, the formalism of Combescot and Nozi\`eres \cite{CN} yields an exact description of the optical response properties. We will restrict in this paper to these approximations and leave the inclusion of electron-electron interactions for future work.

We thus start from the following Hamiltonian:
\begin{eqnarray*}
H = H_M + V_{LM},
\end{eqnarray*}
with
\begin{eqnarray*}
H_M &=& \varepsilon_X \hat{\psi}^\dagger_X\hat{\psi}_X + \sum_k \varepsilon_k \hat{c}^\dagger_k \hat{c}_k + \sum_{k,k^\prime} V^{X-e}_{kk^\prime} \hat{c}^\dagger_{k^\prime}\hat{c}_k\hat{\psi}^\dagger_X\hat{\psi}_X.
\label{matterH}
\end{eqnarray*}
Here, $H_M$ describes the matter degrees of freedom of the quantum well. 
Because the exciton (created by $\hat \psi_X^\dagger$) is assumed to have infinite mass, it is sufficient to consider a single exciton mode that is localized in real space with energy $\varepsilon_X$.
The kinetic energy for the electrons, created with $\hat{c}^\dagger_k$, is given by $\varepsilon_k = k^2$ (units are $\hbar=1, m_e=1/2$). The last term in $H_M$ describes the electron scattering by the exciton. In particular we will use an attractive exciton-electron potential $V^{X-e}<0$, for which it is known that in two dimensions this always results in the presence of a bound state with energy $\varepsilon_T<0$.  Furthermore, we restrict ourselves to the spheric symmetric $l=0$ angular momentum channel. 

The coupling of the QW to the photon field is treated semiclassically as
\begin{eqnarray}
V_{LM}&=&  g A_L e^{-i \omega_L t} \hat{\psi}^\dagger_X+ \textrm{h.c.} 
\label{eq:VLM}
\end{eqnarray}
Here $g$ is the coupling constant between the optical mode in the cavity and the exciton center-off-mass at zero in-plane momentum $k=0$. The amplitude $A_L$ represents a coherent drive by an external laser field.

In order to calculate the optical response function of the QW, we start from the free 2DEG and treat the light-matter coupling term as a perturbation to the system. The optical response $G(t)$ of the QW is then given by linear response theory as \cite{CN}
\begin{eqnarray}
G(t) &=& \langle \textrm{FS}|\mathcal{T}\lbrace\hat{\psi}_X(t)\hat{\psi}^\dagger_X \rbrace|\textrm{FS}\rangle
\label{eq:Gt}
\end{eqnarray}
where $\mathcal{T}$ is the time-ordering operator, and $|\textrm{FS}\rangle$ is the unperturbed Fermi sea (a Slater determinant built with plane waves $|k\rangle$). Using the Heisenberg picture for the time dependence of the operators we can write it as
\begin{eqnarray}
G(t) &=& \langle \textrm{FS}|e^{iHt}\hat{\psi}_Xe^{-iHt}\hat{\psi}^\dagger_X |\textrm{FS}\rangle \crcr
&=& \langle \textrm{FS}|e^{-i(\bar{H}-E_0) t}|\textrm{FS}\rangle .
\label{eq:Gt2}
\end{eqnarray}
The second line expresses that the time evolution of the electrons after the injection of the exciton is governed by the modified Hamiltonian $\bar{H}=H_M(\hat{\psi}^\dagger_X\hat{\psi}_X=1)$, for which $|\textrm{FS}\rangle$ is no longer an eigenstate.
The overlap with the unperturbed Fermi sea will therefore become time-dependent, and in particular for long times it will decay as a powerlaw. The value $E_0=\sum_{k<k_F} \varepsilon_k$ is the ground state energy of $|\textrm{FS}\rangle$. The expectation value \eqref{eq:Gt2} can be straightforwardly computed numerically\cite{CN,Muzyk,knap} as
\begin{eqnarray}
G(t) = \textrm{det}\left[\hat{1}-\hat{n}_F+\hat{n}_F \hat{\lambda}(t)\right].
\label{eq:Gtdet}
\end{eqnarray}
Here, $n_F$ is the Fermi-Dirac distribution. The matrix $\hat\lambda(t)$ is given by
\begin{eqnarray}
\hat\lambda_{kk^\prime}(t)=\sum_p \langle k|p\rangle\langle p|k^\prime \rangle e^{-i(\bar\varepsilon_p-\varepsilon_k)t},
\end{eqnarray}
where one has $\bar H |p\rangle = \bar\varepsilon_p |p\rangle$. Finally the 2DEG absorption is given by
\begin{eqnarray}
\mathcal{A}(\omega)=\frac{1}{\pi}\textrm{Re}\,\int_0^\infty \textrm{d}t\, e^{-i\omega t} \,G(t).
\label{eq:absorption}
\end{eqnarray}

It has been shown\cite{CN} that in the limit $t\gg\varepsilon^{-1}_F$ the function $G(t)$ is given as a sum of two powerlaws,
\begin{equation}
\mathcal{G}(t\gg\varepsilon_F^{-1})  = C_1 \frac{e^{i\omega_1t}}{t^{\alpha_1}}+C_2 \frac{e^{i\omega_2t}}{t^{\alpha_2}}
\label{eq:Fasymp}
\end{equation}
with $C_{1,2}$ some constants. The powerlaw decay is a manifestation of the so-called Anderson orthogonality catastrophe\cite{Anderson} which states that the ground state of the system with and without the scattering potential (due to the exciton) are orthogonal to each other. The powers $\alpha_{1,2}$ can be related to the scattering phase shift at the Fermi level of the conduction electrons scattering off the potential:
\begin{eqnarray}
\alpha_1 &=& \left(\frac{\delta_{F}}{\pi}\right)^2,\crcr
\alpha_2 &=& \left(\frac{\delta_{F}}{\pi}-1\right)^2.
\label{eq:PLexponents}
\end{eqnarray}
The phase shifts  satisfy  $\delta_{F}\in[0,\pi]$, making $\alpha_{1,2}\in[0,1]$. 
Furthermore, since the Laplace transform of the separate terms in \eqref{eq:Fasymp} is given by
\begin{eqnarray}
\int_0^\infty \textrm{d}t\, e^{-i\omega t}\left[ C \frac{e^{i\Omega t}}{t^{\alpha}}\right]=\frac{|C|\Gamma(1-\alpha)e^{-i\frac{\pi\alpha}{2}}}{(\omega-\Omega+i\eta^+)^{1-\alpha}},
\label{eq:SPLaplace}
\end{eqnarray}
with $\Gamma(x)$ the gamma function, it follows that the absorption spectrum contains two singularities, the so-called Fermi edge singularities. The positions of the singularities in frequency domain (`thresholds') is given by\cite{CN}
\begin{eqnarray}
\omega_1 &=& \varepsilon_T+\Delta(\varepsilon_F)\,, \crcr
\omega_2 &=& \Delta(\varepsilon_F)+\varepsilon_F,
\label{eq:thresholds}
\end{eqnarray}
where  $\Delta(\varepsilon_F) = \sum_{p<k_F} (\bar{\varepsilon}_p-\varepsilon_p)$ is the energy difference between the system's ground state before and after the quench. Here, the single particle energies $\bar{\varepsilon}_p$ are obtained by diagonalizing $\bar{H}$.

\begin{figure}[hbtp]
\centering
\includegraphics[scale=0.4]{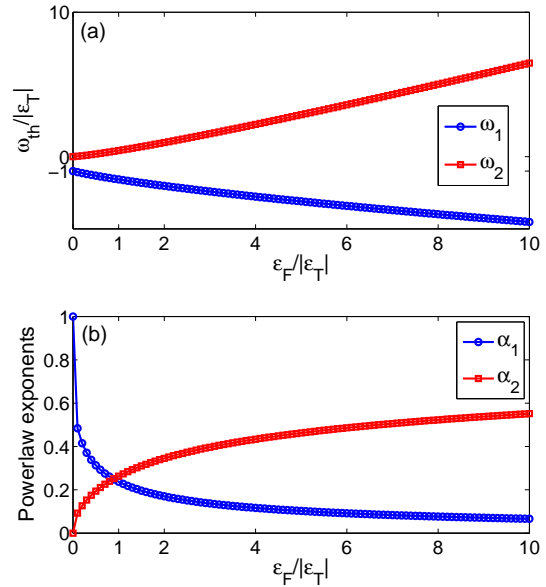}
\caption{a) Thresholds $\omega_{1,2}$ as a function of Fermi energy. b) Powerlaw exponents $\alpha_{1,2}$ as a function of Fermi energy. Both exponents are seen to be smaller than one, corresponding to two singularities in the absorption spectrum. For Fermi energies larger or comparable to the trion binding energy, the trion exponent (blue open circles) starts getting smaller than the exciton exponent (red squares). }
\label{fig:exponentsthresholds}
\end{figure}

Let us now be more specific about the choice for the exciton-electron potential: in the remaining part of this paper we work with an attractive Yukawa potential $V^{X-e}(r)=ae^{-r/b}/r$ with $a<0,b>0$ numerical constants. The thresholds \eqref{eq:thresholds} and the corresponding powerlaw exponents in real time domain \eqref{eq:PLexponents} for the above choice of $V$ are depicted in Fig. \ref{fig:exponentsthresholds} as a function of Fermi energy. Because the scattering potential is attractive, the spectrum $\bar{\varepsilon}_n$ contains a bound state with energy $\varepsilon_T<0$, corresponding to the state in which an electron is bound to the exciton, i.e. a trion.  The trion binding energy is thus given by the amount of energy needed to dissociate the trion into an exciton and an electron. In the following, the binding energy will be used as an energy scale. 

At zero density, the lowest threshold corresponds to the trion and the upper threshold is given by the exciton energy. For sake of simplicity, in the following we will call the lowest threshold the trion, and the highest threshold the exciton for all densities. Still at zero density, we find that the powerlaw exponent in time domain corresponding to the trion ($\alpha_1$) is exactly one, while the exciton exponent $\alpha_2$ is zero. From eq. \eqref{eq:SPLaplace} we then immediately see that the exciton corresponds to a delta function in absorption while the trion is completely flat in frequency domain. For increasing electron density, the trion exponent gets larger than the excitonic exponent, the crossover being for the Fermi energy comparable to the trion binding energy. The latter means that the trion tends to get more delta-function like in absorption for increasing 2DEG density, while the exciton gets flattened out. This behaviour has been numerically verified to hold for several short-range potentials. In particular, we also checked it for the more realistic $1/r^4$ exciton-electron potential. The two thresholds are getting more separated for increasing Fermi energy. This would correspond to an increase of the trion binding energy for higher electron densities. This seems unphysical and should be attributed to the neglect of electron-electron interactions.  

Note that, in order for our model (the exciton as elementary boson) to be valid we must avoid very high electron densities. Specifically, we are restricted to densities satisfying $\varepsilon_F \ll |\varepsilon_X|$. This means that the electron interparticle distance should always be larger than the exciton Bohr radius. Only in this situation it is not possible for the electrons to see the internal structure of the exciton. Typically, in GaAs QW's it holds that $\varepsilon_T \simeq 0.1\varepsilon_X$, so in order for our model to stay valid we must have approximately $\varepsilon_F \leq 10 |\varepsilon_T|$. This explains the horizontal range on the figure \ref{fig:exponentsthresholds}.

\subsection{Lineshapes, oscillator strength, temperature}
\begin{figure}[htbp]
\includegraphics[scale=0.45]{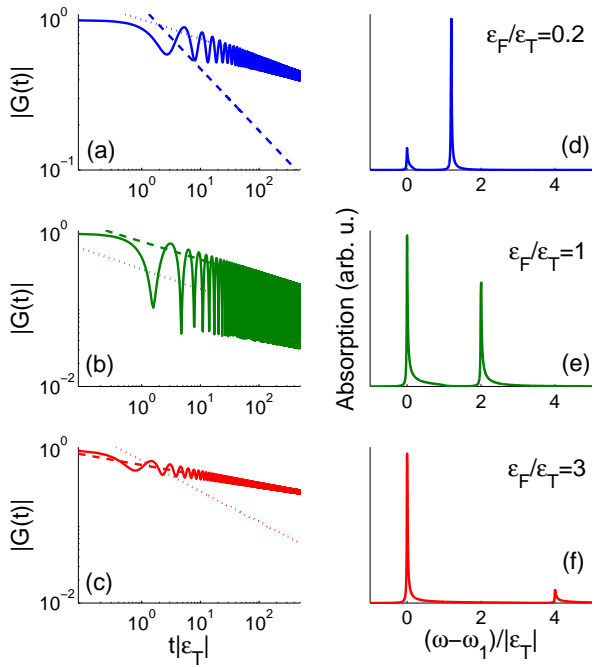}
\caption{a-c) Solid lines: modulus of the response function $|G(t)|$ on a double logaritmic scale for increasing Fermi energy. The dashed (dotted) lines depict the powerlaw decay using the exponents $\alpha_1$ ($\alpha_2$) from eq. \eqref{eq:PLexponents}.  (d-f) Absorption spectra obtained after performing the Laplace transform of the time series. Spectral weight is shifted from the exciton singularity at low densities towards the trion threshold at higher densities.}
\label{fig:TSandLS}
\end{figure}

For the Yukawa potential mentioned earlier, we have numerically computed the response function $G(t)$ in the time domain up to times for which the asymptotic regime, eq. \eqref{eq:Fasymp}, sets in. In Figs. \ref{fig:TSandLS}(a-c) the modulus of the response function is plotted for increasing 2DEG density from top to bottom. Note the double logaritmic scale to evidence the powerlaw nature of the decay. Solid lines correspond to the numerical simulation while the dashed (dotted) line correponds to the powerlaw decay using the trion (exciton) exponent from Fig. \ref{fig:exponentsthresholds}(b). To obtain the absorption, we numerically performed the Laplace transform of the time series using an appropriate damping term to avoid manifestations of the Gibbs phenomenon. 
The corresponding absorption spectra are seen in Figs. \ref{fig:TSandLS}(d-f). All spectra are shifted with respect to the trion threshold $\omega_1$. The spectral lines are seen to have an asymmetric lineshape with a powerlaw tail at the high frequency side of the threshold. This is of course due to the powerlaw exponent in eq. \eqref{eq:SPLaplace}. Only for the real time exponent being exactly zero, the lineshape becomes a symmetric Lorentzian. In theory, the absorption on the left hand side of the trion singularity should be identically zero. Due to the numerical damping, this sudden step is slightly rounded.  Furthermore, for increasing Fermi energy we see that the distance between the two thresholds indeed increases, as mentioned before. The trion is seen to get narrower again for higher densities (the corresponding real time exponent goes to zero for high densities, see Fig. \ref{fig:exponentsthresholds}b). Experimentally many effects such as electron-electron interactions, finite temperature, will nevertheless broaden the spectral line.

\begin{figure}[hbtp]
\includegraphics[scale=0.4]{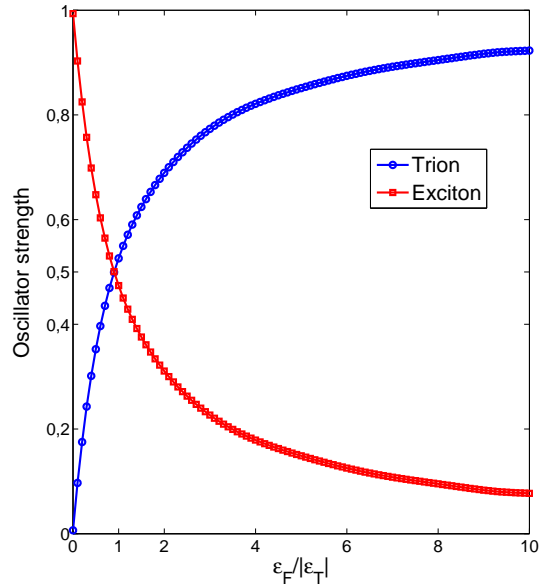}
\caption{Oscillator strength of the trion and exciton as a function of Fermi energy. At zero density the exciton contains all oscillator strength. The trion starts dominating for Fermi energies larger than the trion binding energy. }
\label{fig:OS}
\end{figure}

To quantify which of both singularities is most dominant in the spectrum, we define the oscillator strength as the integral of the absorption spectrum around each singularity. The boundary between the two thresholds is taken as the frequency where the absorption has its minimal value in between. 
Since from the definition \eqref{eq:absorption} we have $\int_{-\infty}^\infty	d\omega A(\omega)=1$,  each oscillator strength is less than 1. Fig. \ref{fig:OS} shows the oscillator strength as a function of Fermi energy.
At zero density all spectral weight is in the exciton (red squares), that is a delta function. This is indeed the only transition that is possible in our model, since at least one electron is needed to form a trion. For increasing Fermi energy, oscillator strength is transferred from the exciton towards the trion (blue circles). Both singularities have equal oscillator strength when the Fermi energy is comparable to the trion binding energy. For very high 2DEG densities most oscillator is gathered in the trion. This is consistent with the spectral lines in Figs. \ref{fig:TSandLS}(d-f) where the exciton is seen to be dominant at low densities, while the trion starts dominating for high electron densities\cite{Perrin}.

\begin{figure}[hbtp]
\includegraphics[scale=0.45]{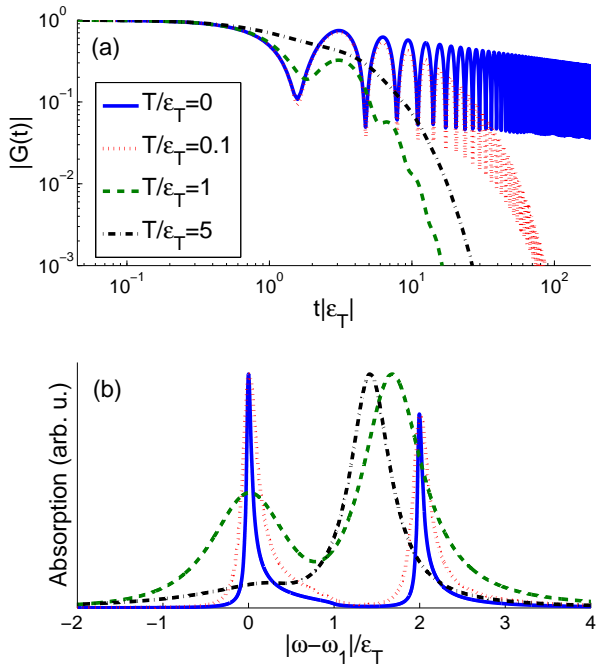}
\caption{a) Time dependence of $|G(t)|$ for a fixed density $\varepsilon_F/\varepsilon_T=1$ and for several temperatures $T/\varepsilon_T=0,\,0.1,\,1,\,5$. For finite $T$, the response function exponentially decays, with a decay time set by the inverse temperature. b) Absorption spectra obtained by Laplace transformation of the time series. Finite temperature causes a broadening of the absorption peaks and makes them more symmetric, i..e Lorentzian.}
\label{fig:LSFiniteTemp}
\end{figure}

Because of the simple form of eq. \eqref{eq:Gtdet}, it is straightforward to investigate finite temperature effects on the spectral lineshapes. In the Fig. \ref{fig:LSFiniteTemp}a we show the time evolution of $G(t)$ for a fixed Fermi energy $\varepsilon_F/\varepsilon_T=1$, and for several temperatures $T/\varepsilon_T$. For finite temperature, the time series are always seen to coincide with the zero temperature result for short times. However, deviations from the $T=0$ start showing up at the time proportional to the inverse temperature, i.e. $t\sim T^{-1}$. For long times the finite temperature series decay exponentially to zero (not visible on the double logaritmic plot). Figure \ref{fig:LSFiniteTemp}b depicts the corresponding absorption spectra. For $T=0$ we have the same lineshape as in Fig. \ref{fig:TSandLS}e. For increasing temperature, the spectral line shapes become broadened and tend to get more Lorentzian. As expected, increasing the temperature too much, causes the two thresholds to merge.

\section{Trion-polariton} \label{sec:TP}

\subsection{Photon spectral function} 

We can now turn on the light-matter interaction between the cavity photon and the charged quantum well. In particular, we are intrested in the linear properties of the coupled light-matter system, i.e. we do not consider photon-photon interactions. The main quantity of intrest is the photon spectral function, given by
\begin{eqnarray*}
D(\omega)=-\textrm{Im}\frac{1}{\omega - \omega_c + i|g_{LM}|^2G(i\omega)}.
\end{eqnarray*}
Here $\omega_c$ is the cavity mode energy and $G(i\omega)$ the Laplace transform of eq. \eqref{eq:Gt}. The microcavity is assumed to be lossless and we only consider a single photon mode at normal incidence.

\begin{figure*}
\centering
\begin{minipage}[t]{.49\textwidth}
\includegraphics[scale=0.45]{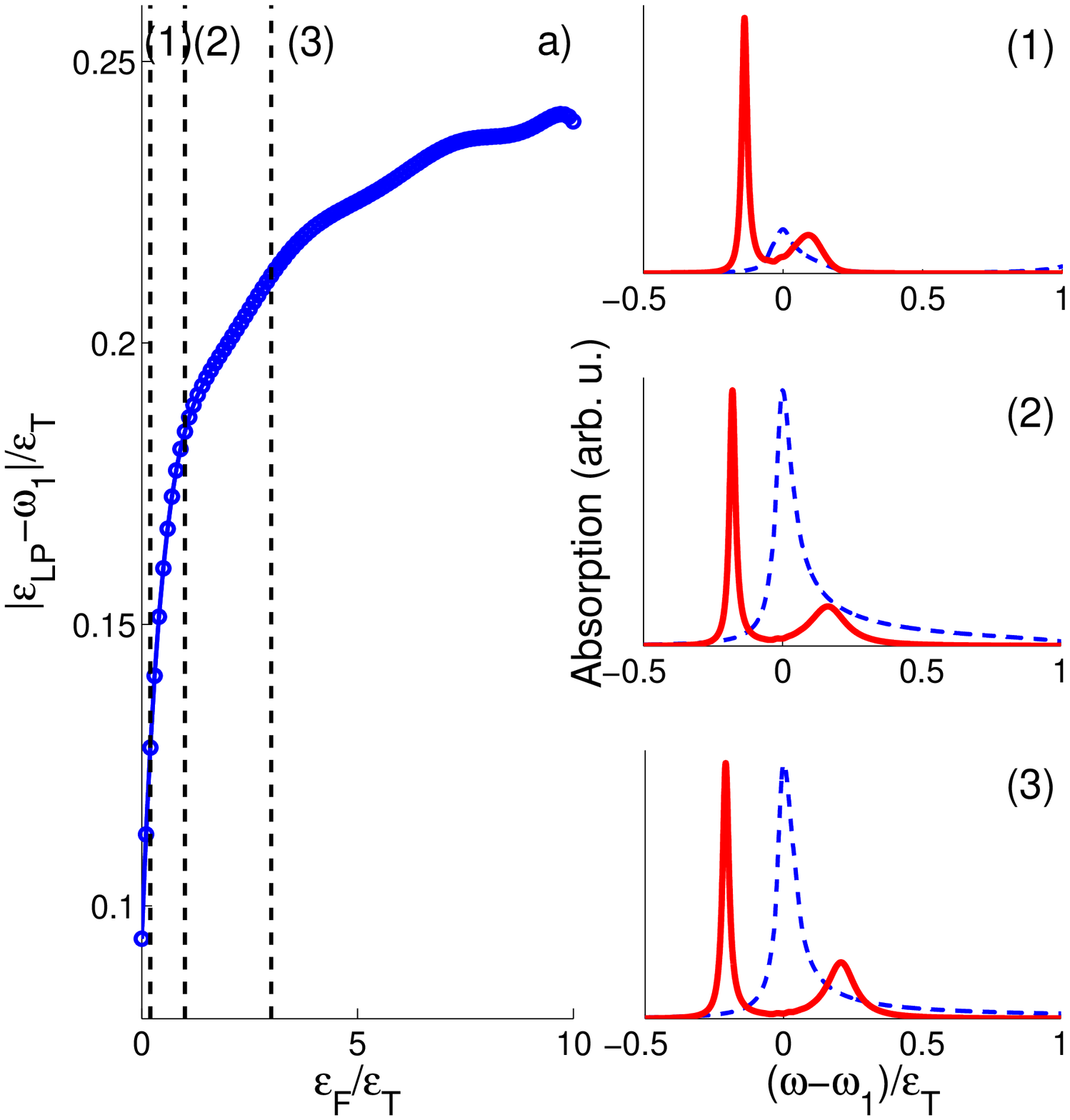}
\caption{a) Lower polariton energy as a function of Fermi energy, obtained by putting the cavity mode into resonance with the trion for all Fermi energies. The light-matter coupling was taken $g_{LM}=0.5|\varepsilon_T|$. (1-3) Polariton lineshapes (red solid) for some fixed Fermi energies (black dashed lines in Fig. a). For reference, the 2DEG absorption is also shown in blue dashed lines. The exciton threshold is not visible because it lies at higher energies.
}\label{fig:Rabismall}
\end{minipage}\hfill
\begin{minipage}[t]{.49\textwidth}
\includegraphics[scale=0.45]{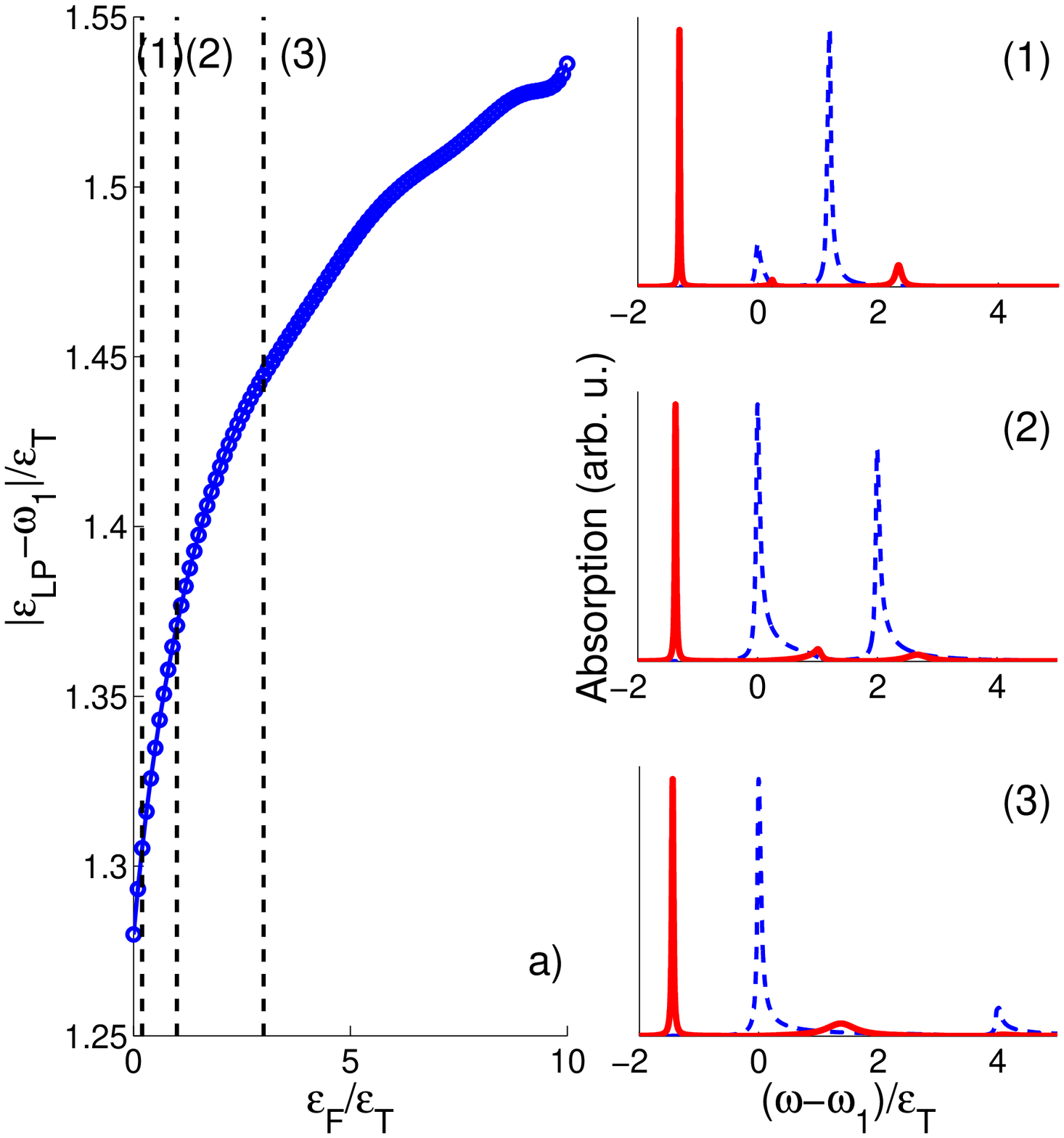}
\caption{Same quantities as in Fig. \ref{fig:Rabismall}, but now for a light-matter coupling $g_{LM}=3|\varepsilon_T|$. The larger light-matter coupling now  admixes the photon with both the trion and the exciton, resulting in an upper polariton with an energy higher than the exciton energy. In (2) the middle polariton is seen to have an asymmetric lineshape, inherited from the 2DEG absorption. }\label{fig:Rabilarge}
\end{minipage}
\end{figure*}

From now on, we restrict our attention to the trion in the 2DEG absorption spectra at zero temperature. In particular, as the trion energy shifts with increasing electron density (see Fig. \ref{fig:exponentsthresholds}a), we can put the photon into resonance with the trion for every density. The resulting lower polariton energy $\varepsilon_{LP}$ is seen in Figs. \ref{fig:Rabismall}a, \ref{fig:Rabilarge}a for two different values of the light-matter coupling, i.e. $g_{LM}= 0.5|\varepsilon_T|$ vs $g_{LM}= 3|\varepsilon_T|$. In both cases, we see a monotonic increase in the trion-polariton `Rabi splitting' as a function of Fermi energy. This can be attributed to the gain of trion oscillator strength as the Fermi energy increases. For low electron density, the increase of the trion oscillator strength can be simply understood in terms of an increased trion-photon overlap \cite{Rapaport}. In recent experiments\cite{atac}, a \emph{decrease} of the Rabi splitting was however observed for higher electron density. Introducing hole recoil and electron-electron interactions could solve this discrepancy.  In the subfigures (1-3) the polariton lineshapes (red solid) are depicted for some fixed Fermi energies, the values being indicated by the black dashed lines in the large figures. For comparison, the 2DEG absorption has also been depicted in blue dashed lines. For small light-matter coupling, the exciton does not influence the cavity mode. The lower polariton now truly corresponds to the trion-polariton as being a superposition of trion and photon. Note that in principle the lower polariton should have zero linewidth, but for visibility we gave it some small value. The other polariton modes intrinsically possess a finite linewidth due to the finite absorption of the 2DEG at the corresponding polariton energies. For larger light-matter coupling, the photon starts admixing with the exciton resulting in a non-zero quasiparticle shift at the lowest electron density under consideration. We also see a clearly visible polariton mode above the exciton threshold. Furthermore, in Fig. \ref{fig:Rabilarge}(2) the middle polariton is seen to have an asymmetric lineshape with a large tail towards the low frequencies. This feature stems from the interplay between the two skewed lineshapes from both the trion and exciton and the fact that they both contribute equally to the polariton formation because of their oscillator strengths being the same at the corresponding Fermi energy. 

\subsection{Electron density in the trion-polariton state} 

One of the attractive features of microcavity polaritons is their strong optical nonlinearity, allowing the creation of an interacting polariton quantum `fluid'. A Feshbach resonance mechanism for enhancing the polariton interactions \cite{Vladimirova} has recently been investigated\cite{takemura}. Unfortunately, in the current state of the art microcavities, the single photon nonlinearities are too weak in order to enter the photon blockade regime. 

It has been recently suggested that the electrons in the QW could enhance interactions between the polaritons \cite{atac}. In the previous part of the paper, we have shown the lower trion-polariton to be the good quasi particle for a highly doped QW embedded in a planar microcavity. Because the matter component of the polaritons is responsible for the interactions, it is instructive to investigate the spatial structure of the trion-polariton. In particular, we will calculate the electron density in the lower polariton state through the electron-exciton density correlation function:

\begin{eqnarray}
n_e(r)&=&\frac{\langle \hat{n}_e(r)\hat{n}_X(0) \rangle}{\langle\hat{n}_X(0) \rangle},
\label{eq:densdens}
\end{eqnarray} 
with $\hat{n}(x)=\hat{\psi}^\dagger(x)\hat{\psi}(x)$ the density operator at position $x$.
Because the exciton is assumed to be infinitely heavy, we can place it in the origin without loss of generality. Since we only consider s-wave scattering, the meaning of the spatial coordinate $r$ is the relative radial distance from the electrons to the exciton center-of-mass.
We start again from the unperturbed Fermi sea and consider the Hamiltonian $V_{LM}$ given by Eq. \eqref{eq:VLM} as a small perturbation to the 2DEG. Using Kubo's response theory, we find that the lowest order contribution to the above correlation function is of second order in the perturbation, i.e. proportional to $|gA_L|^2$. It is given by
\begin{widetext}
\begin{eqnarray}
\langle \hat{n}_e(r)\hat{n}_X(0) \rangle &=& |gA_L|^2\lim_{\eta\to 0^+}\lim_{t_0\to -\infty}\int_{t_0}^tds\,e^{\eta s}\int_{t_0}^sd\tau \,e^{\eta \tau}e^{i\omega_L(s-\tau)}\,\langle \textrm{FS}| \hat{\psi}_X(0,s) \,[\hat{n}_e(r,t)\hat{n}_X(0,t)]\, \hat{\psi}^\dagger_X(0,\tau)|\textrm{FS}\rangle+\textrm{h.c.}\nonumber \\
&=&  |gA_L|^2\sum_{n,m}\phi_n(r)\phi^*_m(r)\int_{-\infty}^\infty d\omega\, g_{nm}(\omega)\frac{1}{\omega-(\omega_L-\bar{\varepsilon}_n)}\frac{1}{\omega-(\omega_L-\bar{\varepsilon}_m)}.
\label{eq:xedens}
\end{eqnarray}
\end{widetext}
Here, we adiabatically turned on the laser in order to find the stationary state of the electron density under the continous wave excitation (see details in the appendix). The wave functions $\phi_n(r)$ are the single particle eigenstates of the Hamiltonian $\bar{H}$ in position space, satisfying $\bar{H} \phi_n(r)=\bar{\varepsilon}_n\phi_n(r)$. The matrix $g_{nm}(\omega)$ is given as 
\begin{eqnarray}
g_{nm}(\omega)&=&\frac{1}{\pi}\textrm{Re}\,\int_0^\infty \textrm{d}t\, e^{-i\omega t} g_{nm}(t),\nonumber\\
g_{nm}(t)&=& \sum_{k,q} \phi_n^*(k)\phi_m(q)\langle \textrm{FS}|\hat{c}^\dagger_qe^{-i(\bar{H}-E_0)t}\hat{c}_k|\textrm{FS}\rangle.
\label{eq:minors}
\end{eqnarray}
The expectation value in the above equation looks similar to eq. \eqref{eq:Gt2} and it can again be calculated numerically \cite{CN}. This means that our results on the electron density properly take into account the Anderson orthogonality catastrophe and the Fermi edge singularity physics.

The expectation value \eqref{eq:densdens} should be calculated in the lower polariton state. In the resulting expression eq. \eqref{eq:xedens} the polariton shows up through the laser frequency $\omega_L$. In particular, we consider resonant continous wave excitation, thus with the frequency of the laser resonant with the lower polariton energy $\varepsilon_{LP}$. Note that $\varepsilon_{LP}$ is a free parameter in this calculation, but its actual value has been calculated in the previous part of this paper, see Figs. \ref{fig:Rabismall}a, \ref{fig:Rabilarge}a.\newline
The exciton density in the lower polariton state can be calculated analogously. It is found to be \begin{eqnarray}
\langle\hat{n}_X(0) \rangle = |gA_L|^2\int_{-\infty}^\infty d\omega\, \frac{\mathcal{A}(\omega)}{(\omega-\omega_L)^2}
\end{eqnarray}
where the absorption $\mathcal{A}(\omega)$ is given in eq. \eqref{eq:absorption}.

\begin{figure}[htbp]
\includegraphics[scale=0.45]{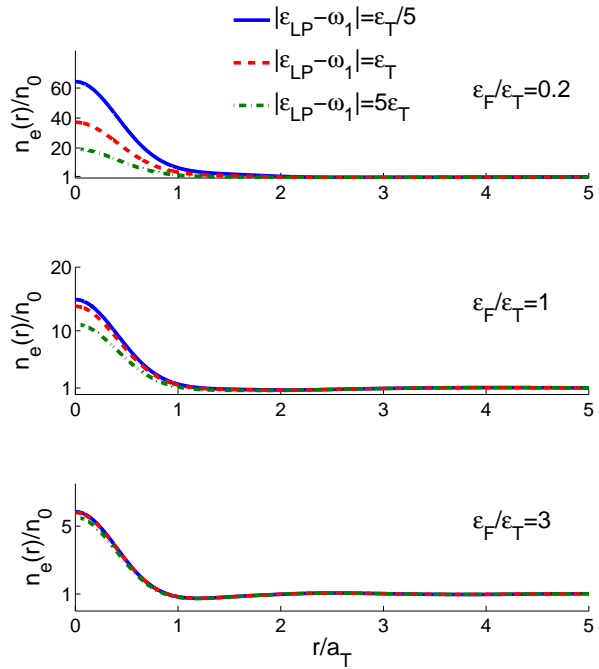}
\caption{Electron density in the lower polariton state (normalized with respect to the unperturbed density $n_0=k_F^2/4\pi$) as a function of radial distance to the exciton center-of-mass. At large distances, the density stays unperturbed as the correlation function always asymptotically goes to $1$. For a high density 2DEG the electron density does not depend too much on the Rabi frequency because the electrons screen the exciton on a time scale of the inverse Fermi energy. For smaller Fermi energy, a low Rabi frequency still allows the electrons to screen the exciton. For higher Rabi frequency, the electrons do not have enough time to react onto the presence of the exciton so that the electron density is unperturbed.  }
\label{fig:g2}
\end{figure}

The electron density along the radial direction and in the lower polariton state is shown in Fig. \ref{fig:g2} for several Fermi energies and polariton energies. The trion Bohr radius $a_T$ is taken as the lengthscale corresponding to the bound state $\psi(r)\sim \exp(-r/a_T)$ with energy $\varepsilon_T$.

For the lowest Fermi energy, a small Rabi frequency (blue solid lines), defined as the energy difference between the lower polariton and the trion threshold $\omega_1$, is seen to increase the electron density at short distances from the exciton. This can be understood as follows: when the photon is annihilated and an exciton is created, the Fermi sea is shaken up due to the appearance of the exciton. For small Rabi frequency, it takes longer time for the exciton to be reconverted into a photon. This means that there is a longer time for the electrons in the Fermi sea to adjust themselves to the presence of the exciton: a screening cloud of electrons is able to build up near the exciton. The reason for the increase of electron density is the attractive exciton-electron potential, i.e. the presence of the bound state. For larger Rabi frequency, there will be much less time for the electrons to reorganize; as soon as they start readjusting, the exciton has already dissapeared and turned into a photon. This is seen by the green dashed-dotted line where the exciton density tends to flatten. For large Fermi energies (comparable to or larger than the trion binding energy), the electron density does not depend too much on the Rabi frequency. We argue that the time it takes for the individual electrons to adjust to the presence of the scattering potential, is proportional to the inverse Fermi energy. The larger the Fermi energy, the less time it takes to reorganize the electrons at the Fermi level and the system will always be able to build up a screening cloud. Nevertheless, the absolute value of $n_e(r)/n_0$ decreases for increasing Fermi energy, simply because there is less time to screen the exciton. In all cases, the electron density oscillates and the envelope decays as $1/r^2$ at large distances, corresponding to the Friedel oscillations.

From the above, we see that there should be a frequency window in which the trion gets screened on a length scale comparable to the bare trion Bohr radius, making it a truly localized, electrically charged particle, while still ensuring strong coupling. The Rabi frequency should be small enough such that there is time for the electrons to form for the trion, by binding to the exciton. Since the formation time typically goes as $1/\varepsilon_T$, one should maintain $\Omega_R<\varepsilon_T$. On the other hand, one should avoid running into the weak coupling regime by taking the Rabi frequency too small.  

\section{Conclusion and outlook}\label{ConclusionAndOutlook}
We have set up an effective model describing the formation of trion-polaritons in a microcavity embedding a doped quantum well. Taking into account important many-body effects such as the Anderson orthogonality catastrophe and Fermi edge singularity, we were able to qualitatively describe the 2DEG absorption, the crossover of oscillator strength from exciton to trion and the polariton lineshapes. The role of finite temperature has been discussed as well. Finally, we calculated the electron density profile in the lower polariton state and concluded that for a specific range of Rabi frequencies, the trion-polariton behaves as an electrically charged localized particle.

A major simplification in our model was the neglect of electron-electron interactions. Including this physics would probably solve the decreasing linewidth of the trion for increasing electron density. Also, the binding energy of the trion then becomes a function of Fermi energy instead of being a fixed parameter. Most importantly, a self-consistent treatment of electron-electron interactions in the calculation of the electron density profile is needed to see whether a charge build-up near the exciton is still possible.  

\section*{Acknowledgements}
We are greatful to Ata\c{c} Imamoglu, Cristiano Ciuti, Charles Grenier and Dries Sels for stimulating discussions. This work was financially supported by the UA-LP program.

\appendix 
\onecolumngrid

\section{Exciton-electron density correlation function}
In this appendix we derive, using Kubo's reponse theory, the expression for the electron-exciton density correlation function.\newline
Starting from the unperturbed Fermi sea and treating the Hamiltonian \eqref{eq:VLM} as a perturbation to the system, we need to go second order in this perturbation to have a first non-zero contribution to the density-density correlation function. Using standard Kubo response theory we find

\begin{eqnarray}
\langle \hat{n}_{e}\left( \mathbf{r,}t\right) \hat{%
n}_{X}\left( \mathbf{0,}t\right) \rangle =\left\vert gA_{L}\right\vert ^{2}\int_{t_0 }^{t}ds\int_{t_0 }^{s}d\tau \,\,\left[ e^{i\omega
_{L}\left( s-\tau \right) }\langle\textrm{FS}| \,\hat{\psi}_{X}\left( \mathbf{0,}s
\right) \hat{n}_{e}\left({\bf r}, t\right) \hat{n}_{X}({\bf 0},t)\,\hat{\psi}
_{X}^{\dagger }\left( \mathbf{0,}\tau\right)|\textrm{FS} \rangle + \text{h.c.}\right]
\label{eq:elex}
\end{eqnarray}
where the remaining expectation values need to be calculated in the unperturbed state.\newline
We perform the calculation in two dimensions. Restriction to the spheric symmetric states comes down to using the proper Hankel transform pairs. Using the interaction picture for the time evolution of the operators we obtain
\begin{eqnarray*}
\langle\textrm{FS}| \,\hat{\psi}_{X}\left( \mathbf{0,}s
\right) \hat{n}_{e}\left({\bf r}, t\right) \hat{n}_{X}({\bf 0},t) \,\hat{\psi}
_{X}^{\dagger }\left( \mathbf{0,}\tau\right)|\textrm{FS} \rangle
&=&e^{iE_{0}\left( s-\tau \right) }\langle \textrm{FS}|e^{-i\bar{H}s }\psi
_{e}^{\dagger }\left( \mathbf{r},t\right) \psi _{e}\left( \mathbf{r},t\right)
e^{i\bar{H}\tau}|\textrm{FS}\rangle ,
\end{eqnarray*}
where $\bar{H}$ is the Hamiltonian governing the dynamics of the electrons in presence of the exciton scattering potential and $E_0$ is the ground state energy of the Fermi sea without the exciton.\newline
Using the Fourier transform for the system enclosed in a box of area $S$, we obtain
\begin{eqnarray}
\langle \textrm{FS}|e^{-i\bar{H}s }\psi
_{e}^{\dagger }\left( \mathbf{r},t\right) \psi _{e}\left( \mathbf{r},t\right)
e^{i\bar{H}\tau}|\textrm{FS}\rangle 
&=&\frac{1}{S}\sum_{\mathbf{k,q}}e^{i\left( \mathbf{k-q}\right) \mathbf{r}%
}\,\langle \textrm{FS}|e^{-i\bar Hs}c_{\mathbf{k}}^{\dagger }\left( t\right) c_{%
\mathbf{q}}\left( t\right) \,e^{i\bar H\tau }|\textrm{FS}\rangle ,
\end{eqnarray}
where the plane wave creation operators satisfy $\left\lbrace c_{\mathbf{k}},c_{\mathbf{q}}^{\dagger }\right\rbrace=\delta_{{\bf k,q}}$. The time evolution of the operators is governed by the Hamiltonian in presence of the exciton:
\begin{eqnarray}
\bar H &=&  \sum_k \varepsilon_k \hat{c}^\dagger_k \hat{c}_k + \sum_{k,k^\prime} V^{X-e}_{kk^\prime} \hat{c}^\dagger_{k^\prime}\hat{c}_k.
\end{eqnarray}
The eigenstates are no longer plane waves and we therefore make a unitary transformation from the plane wave basis to the basis of eigenstates of $\bar H$, we call them scattering states. The latter are characterized by their principal quantum number $n$ and the angular momentum  $l$, because the scattering potential under consideration is spheric symmetric. This gives
 \begin{eqnarray}
&&\frac{1}{S}\sum_{\mathbf{k,q}}e^{i\left( \mathbf{k-q}\right) \mathbf{r}%
}\,\langle \textrm{FS}|e^{-i\bar Hs}c_{\mathbf{k}}^{\dagger }\left( t\right) c_{%
\mathbf{q}}\left( t\right) \,e^{i\bar H\tau }|\textrm{FS}\rangle \\
&=&\frac{1}{S}\sum_{\mathbf{k,q}}e^{i\left( \mathbf{k-q}\right) \mathbf{r}%
}\,\sum_{n,m=0}^{\infty }\sum_{l,p=-\infty }^{\infty }\langle \mathbf{q}%
|n,l\rangle \langle m,p%
|\mathbf{k}\rangle \,\langle \textrm{FS}|e^{-i\bar Hs}c_{mp}^{\dagger }\left( t\right) \hat{c}_{nl}\left( t\right) \,e^{i\bar H\tau }|\textrm{FS}\rangle .
\end{eqnarray}
Since the Hamiltonian $\bar H$ is diagonal in the operators $c_{nl}^{\dagger }$ we have a trivial time evolution. We than have

\begin{eqnarray}
&&\frac{1}{S}\sum_{\mathbf{k,q}}e^{i\left( \mathbf{k-q}\right) \mathbf{r}%
}\,\sum_{n,m=0}^{\infty }\sum_{l,p=-\infty }^{\infty }\langle \mathbf{q}%
|n,l\rangle \langle m,p%
|\mathbf{k}\rangle \,\langle \textrm{FS}|e^{-i\bar Hs}c_{mp}^{\dagger }\left( t\right) \hat{c}_{nl}\left( t\right) \,e^{i\bar H\tau }|\textrm{FS}\rangle 
\\
&=&\frac{1}{S}\sum_{\mathbf{k,q}}e^{i\left( \mathbf{k-q}\right) \mathbf{r}%
}\sum_{n,m=0}^{\infty }e^{i\bar{\varepsilon}_{m}t}e^{-i\bar{\varepsilon}%
_{n}t}\sum_{l,p=-\infty }^{\infty }\langle \mathbf{q}|n,l\rangle \,\langle
m,p|\mathbf{k}\rangle \,\langle \textrm{FS}|e^{-i\bar Hs}\hat{c}_{mp}^{\dagger
}\,\hat{c}_{nl}\,e^{i\bar H\tau }|\textrm{FS}\rangle  \\
&=&\frac{1}{S}\sum_{\mathbf{k,q}}e^{i\left( \mathbf{k-q}\right) \mathbf{r}%
}\sum_{n,m=0}^{\infty }e^{i\bar{\varepsilon}_{m}t}e^{-i\bar{\varepsilon}%
_{n}t}\sum_{l,p=-\infty }^{\infty }\langle \mathbf{q}|n,l\rangle \,\langle
m,p|\mathbf{k}\rangle \,\langle \textrm{FS}|e^{-iH_{f}s}\hat{c}_{mp}^{\dagger
}e^{i\bar Hs}e^{-i\bar H\left( s-\tau \right) }e^{-i\bar H\tau }\,\hat{c}%
_{nl}\,e^{i\bar H\tau }|\textrm{FS}\rangle  \\
&=&\frac{1}{S}\sum_{\mathbf{k,q}}e^{i\left( \mathbf{k-q}\right) \mathbf{r}%
}\sum_{n,m=0}^{\infty }e^{i\bar{\varepsilon}_{m}t}e^{-i\bar{\varepsilon}%
_{n}t}\sum_{l,p=-\infty }^{\infty }\langle \mathbf{q}|n,l\rangle \,\langle
m,p|\mathbf{k}\rangle \,\langle \textrm{FS}|\hat{c}_{mp}^{\dagger }\left(
-s\right) e^{-i\bar H\left( s-\tau \right) }\hat{c}_{nl}\left( -\tau \right)
|\textrm{FS}\rangle  \\
&=&\frac{1}{S}\sum_{\mathbf{k,q}}e^{i\left( \mathbf{k-q}\right) \mathbf{r}%
}\sum_{n,m=0}^{\infty }e^{i\bar{\varepsilon}_{m}t}e^{-i\bar{\varepsilon}%
_{n}t}e^{-i\bar{\varepsilon}_{m}s}e^{i\bar{\varepsilon}_{n}\tau
}\sum_{l,p=-\infty }^{\infty }\langle \mathbf{q}|n,l\rangle \,\langle m,p|%
\mathbf{k}\rangle \,\langle \textrm{FS}|\hat{c}_{mp}^{\dagger
}e^{-i\bar H\left( s-\tau \right) }\hat{c}_{nl}|\textrm{FS}\rangle ,
\end{eqnarray}
where in the third line we introduced the unit operator. The initial state $|\textrm{FS}\rangle$ is built with plane waves but the operator $\hat c_{nl}$ destroys a scattering state. We therefore make again a transformation from scattering states to plane wave states, finally yielding
\begin{eqnarray}
&&\langle \textrm{FS}|e^{-i\bar{H}s }\psi
_{e}^{\dagger }\left( \mathbf{r},t\right) \psi _{e}\left( \mathbf{r},t\right)
e^{i\bar{H}\tau}|\textrm{FS}\rangle \\
&=&\frac{1}{S}\sum_{\mathbf{k,q}}e^{i\left( \mathbf{k-q}\right) \mathbf{r}%
}\sum_{n,m=0}^{\infty }e^{i\bar{\varepsilon}_{m}t}e^{-i\bar{\varepsilon}%
_{n}t}e^{-i\bar{\varepsilon}_{m}s}e^{i\bar{\varepsilon}_{n}\tau
}\sum_{l,p=-\infty }^{\infty }\langle \mathbf{q}|n,l\rangle \,\langle m,p|%
\mathbf{k}\rangle \, \sum_{\mathbf{K,Q}}\mathbf{\,\langle }n,l\mathbf{|K\rangle }\langle{\mathbf{Q}|m,p}\rangle\langle\textrm{FS}|\hat{c}_{\mathbf{Q}}^{\dagger
}e^{-i\bar H\left( s-\tau \right) }\hat{c}_{\mathbf{K}}|\textrm{FS}\rangle. \nonumber
\end{eqnarray}
The sum over $\mathbf{k,q}$ defines the wave function in position space. Writing $\langle \mathbf{k}|n,l\rangle=\phi_{nl}(\mathbf{k})$, the eigenstates of the Hamiltonian $\bar H$ in plane wave basis, we define
\begin{eqnarray*}
\phi_{nl}(\mathbf{r})=\frac{1}{\sqrt{S}}\sum_{\mathbf{k}}e^{-i \mathbf{k} \mathbf{r}}\,\phi_{nl}(\mathbf{k}),
\end{eqnarray*}
and we have
\begin{eqnarray}
&&\langle\textrm{FS}| \,\hat{\psi}_{X}\left( \mathbf{0,}s
\right) \hat{n}_{e}\left({\bf r}, t\right) \hat{n}_{X}({\bf 0},t) \,\hat{\psi}
_{X}^{\dagger }\left( \mathbf{0,}\tau\right)|\textrm{FS} \rangle \\
&=&e^{iE_{0}\left( s-\tau \right) }\langle \textrm{FS}|e^{-i\bar{H}s }\psi
_{e}^{\dagger }\left( \mathbf{r},t\right) \psi _{e}\left( \mathbf{r},t\right)
e^{i\bar{H}\tau}|\textrm{FS}\rangle \\
&=&\sum_{n,m=0}^{\infty }\sum_{l,p=-\infty }^{\infty }\phi_{nl}(\mathbf{r})\phi^*_{mp}(\mathbf{r})\,e^{i\bar{\varepsilon}_{m}t}e^{-i\bar{\varepsilon}%
_{n}t}e^{-i\bar{\varepsilon}_{m}s}e^{i\bar{\varepsilon}_{n}\tau
} \, \sum_{\mathbf{K,Q}} \phi^*_{nl}(\mathbf{K})\phi_{mp}(\mathbf{Q}) \langle\textrm{FS}|\hat{c}_{\mathbf{Q}}^{\dagger
}e^{-i(\bar H-E_0)\left( s-\tau \right) }\hat{c}_{\mathbf{K}}|\textrm{FS}\rangle \\
&=&\sum_{n,m=0}^{\infty }\sum_{l,p=-\infty }^{\infty }\phi_{nl}(\mathbf{r})\phi^*_{mp}(\mathbf{r})\,e^{i\bar{\varepsilon}_{m}t}e^{-i\bar{\varepsilon}%
_{n}t}e^{-i\bar{\varepsilon}_{m}s}e^{i\bar{\varepsilon}_{n}\tau
} \, g_{nm}(s-\tau).
\end{eqnarray}
Here, we introduced shorthand notation for the double sum over $\mathbf{K,Q}$ (dropping the angular momentum indices in order not to overload the notation):
\begin{eqnarray}
g_{nm}(t)=\sum_{\mathbf{K,Q}} \phi^*_{nl}(\mathbf{K})\phi_{mp}(\mathbf{Q}) \langle\textrm{FS}|\hat{c}_{\mathbf{Q}}^{\dagger
}e^{-i(\bar H-E_0)t }\hat{c}_{\mathbf{K}}|\textrm{FS}\rangle.
\end{eqnarray}
This expression is exactly the one from eq. \eqref{eq:minors} in the paper. The above expectation value should be plugged into equation \eqref{eq:elex}. This gives
\begin{eqnarray}
&&\langle \hat{n}_{e}\left( \mathbf{r,}t\right) \hat{%
n}_{X}\left( \mathbf{0,}t\right) \rangle \\
&=&\left\vert gA_{L}\right\vert ^{2}\sum_{n,m=0}^{\infty }\sum_{l,p=-\infty }^{\infty }\phi_{nl}(\mathbf{r})\phi^*_{mp}(\mathbf{r})\,e^{i\bar{\varepsilon}_{m}t}e^{-i\bar{\varepsilon}%
_{n}t} \int_{t_0 }^{t}ds\int_{t_0 }^{s}d\tau \,\, e^{i\omega
_{L}\left( s-\tau \right) } e^{-i\bar{\varepsilon}_{m}s}e^{i\bar{\varepsilon}_{n}\tau
} \, g_{nm}(s-\tau) + \text{h.c.}
\end{eqnarray}
The hermitian conjugate term can be shown to be the same as interchanging $s$ and $\tau$, yielding
\begin{eqnarray}
&&\langle \hat{n}_{e}\left( \mathbf{r,}t\right) \hat{%
n}_{X}\left( \mathbf{0,}t\right) \rangle \\
&=&\left\vert gA_{L}\right\vert ^{2}\sum_{n,m=0}^{\infty }\sum_{l,p=-\infty }^{\infty }\phi_{nl}(\mathbf{r})\phi^*_{mp}(\mathbf{r})\,e^{i\bar{\varepsilon}_{m}t}e^{-i\bar{\varepsilon}%
_{n}t} \int_{t_0 }^{t}ds\int_{t_0 }^{t}d\tau \,\, e^{i\omega
_{L}\left(s- \tau \right) } e^{-i\bar{\varepsilon}_{m}s}e^{i\bar{\varepsilon}_{n}\tau
} \, g_{nm}(s-\tau),
\end{eqnarray}
where now we can both have $s<\tau$ and $s>\tau$. Introducing the Fourier transformation of $g_{nm}$, we can now easily do the integration over $s,\,\tau$. Furthermore, in order to obtain the stationary state under the continous excitation, we add a small exponential term, meaning we slowly (for example, as compared to the lifetime of the cavity mode) turn on the laser. If we then take $t_0\rightarrow -\infty$, this will eliminate all transient behaviour, originating from the initial time $t_0$. In other words, we do not want our result to depend on $t_0$.\newline
So, introducing the Fourier transform for a function $\varrho(t)$,
\begin{eqnarray*}
\rho \left( t\right) &=&\int_{-\infty }^{\infty }\,d\omega \,e^{-i\omega
t}\,\varrho \left( \omega \right) \\
\varrho \left( \omega \right) &=&\frac{1}{2\pi }\int_{-\infty }^{\infty
}\,dt\,e^{i\omega t}\,\rho \left( t\right),
\end{eqnarray*}%
we obtain, by adiabatically turning on the laser (only focus on the time integrals)
\begin{eqnarray*}
&& e^{i\bar{\varepsilon}_{m}t}e^{-i\bar{\varepsilon}%
_{n}t}\lim_{\eta\to 0^+}\lim_{t_0 \to -\infty}\int_{t_0 }^{t}ds\, e^{\eta^+s}\int_{t_0 }^{t}d\tau \,e^{\eta^+\tau}\, e^{i\omega
_{L}\left( s-\tau \right) } e^{-i\bar{\varepsilon}_{m}s}e^{i\bar{\varepsilon}_{n}\tau
} \, \int_{-\infty }^{\infty }\,d\omega \,e^{-i\omega
(s-\tau)}\,g_{nm}\left( \omega \right)\\
&=&e^{i\bar{\varepsilon}_{m}t}e^{-i\bar{\varepsilon}%
_{n}t}\lim_{\eta\to 0^+}\lim_{t_0 \to -\infty}\int_{-\infty }^{\infty }\,d\omega \,g_{nm}\left( \omega \right) \int_{t_0 }^{t}ds\, e^{\eta^+s}\int_{t_0 }^{t}d\tau \,e^{\eta^+\tau}\, e^{i\omega
_{L}\left( s-\tau \right) } e^{-i\bar{\varepsilon}_{m}s}e^{i\bar{\varepsilon}_{n}\tau
} \, \,e^{-i\omega
(s-\tau)}\\
&=&\int_{-\infty }^{\infty }\,d\omega \,g_{nm}\left( \omega \right)\frac{1}{\omega-(\omega_L-\bar{\varepsilon}_m)}\frac{1}{\omega-(\omega_L-\bar{\varepsilon}_n)}.
\end{eqnarray*}%
The last line follows from straightforward integration and taking the proper limits. Restoring the double sum over $n,m$, we obtain equation \eqref{eq:xedens} in the article. The exciton density can be calculated completely analogously.


\begin{thebibliography}{99}

\bibitem{iac_review} I. Carusotto, and C. Ciuti, Rev. Mod. Phys. {\bf 85}, 299-366 (2013).
\bibitem{laussy} F. P. Laussy, A. V. Kavokin, and I. A. Shelykh, Phys. Rev. Lett. {\bf 104}, 106402 (2010).
\bibitem{GabbayCohen} A. Gabbay, Y. Preezant, E. Cohen, B. M. Ashkinadze, and L. N. Pfeiffer, Phys. Rev. Lett. {\bf 99}, 157402 (2007).

\bibitem{atac}  S. Smolka, W. Wuester, F. Haupt, S. Faelt, W. Wegscheider, and A. Imamoglu, Science {\bf 346}, 6207 (2014).

\bibitem{Lagoudakis} P. G. Lagoudakis, M. D. Martin, J. J. Baumberg, A. Qarry, E. Cohen, and L. N. Pfeiffer, Phys. Rev. Lett. {\bf 90}, 206401 (2003).
\bibitem{Perrin} M. Perrin, P. Senellart, A. Lema\^itre, and J. Bloch, Phys. Rev. B {\bf 72}, 075340 (2005).
\bibitem{Das} A. Das, B. Xiao, S. Bhowmick, and P. Bhattacharya, Applied Physics Letters {\bf 101}, 131112 (2012).
\bibitem{Kavokin} G. Malpuech, A. Kavokin, A. Di Carlo, and J. J. Baumberg, Phys. Rev. B {\bf 65}, 153310 (2002).
\bibitem{mahan} G. D. Mahan, Phys. Rev. {\bf 153}, 882 (1967).
\bibitem{nozieres} P. Nozi\`eres, and C.T. De Dominicis, Phys. Rev. {\bf 178}, 1097 (1969).
\bibitem{Anderson} P. Anderson, Phys. Rev. Lett. {\bf 18}, 1049 (1967).
\bibitem{knap} M. Knap, A. Shashi, Y. Nishida, A. Imambekov, D. A. Abanin, and E. Demler, Phys. Rev. X {\bf 2}, 041020 (2012).
\bibitem{demler} G. Refael, and E. Demler, Phys. Rev. B {\bf 77}, 144511 (2008).
\bibitem{Schirotzek} A. Schirotzek, C. H. Wu, A. Sommer, and M. W. Zwierlein, Phys. Rev. Lett. {\bf 102}, 230402 (2009).
\bibitem{glazov} N. S. Averkiev, and M. M. Glazov, Phys. Rev. B {\bf 76}, 045320 (2007).
\bibitem{Averkiev} N. S. Averkiev, M. M. Glazov, and M. M. Voronov, Solid. State. Comm. {\bf 152}, 395 (2012).
\bibitem{BW} M. Baeten, and M. Wouters, Phys. Rev. B, {\bf 89}, 245301 (2014).
\bibitem{Rapaport} R. Rapaport, R. Harel, E. Cohen, Arza Ron, E. Linder, and L. N.
Pfeiffer, Phys. Rev. Lett. {\bf 84}, 1607 (2000).


\bibitem{Bloch} D. Bajoni, M. Perrin, P. Senellart, A. Lema\^itre, B. Sermage, and J. Bloch, Phys. Rev. B {\bf 73}, 205344 (2006).

\bibitem{CN} M. Combescot, and P. Nozi\` eres, Le Journal de Physique, {\bf 32}, 913 (1971).
\bibitem{Muzyk} N. d'Ambrumenil, and B. Muzykantskii, Phys. Rev. B {\bf 71}, 045326 (2005).

\bibitem{Vladimirova} M. Vladimirova, S. Cronenberger, D. Scalbert, K. V. Kavokin, A. Miard, A. Lema\^itre, J. Bloch, D. Solnyshkov, G. Malpuech, and A. V. Kavokin, Phys. Rev. B {\bf 82}, 075301, (2010). 

\bibitem{takemura} N. Takemura, S. Trebaol, M. Wouters, M. T. Portella-Oberli and B. Deveaud, Nat. Phys. {\bf 10}, 500 (2014) .














\end{thebibliography}
\end{document}